\documentclass[aps,prd,twocolumn,showpacs]{revtex4-1}
\usepackage{graphicx}
\usepackage{subfigure}
\usepackage{multirow}
\usepackage{hyperref} 

\newcommand{\hpp}{H^{\pm\pm}}
\def\dzsig  {\ensuremath{|d_0|/\sigma(d_0)}} 
\def\ifb{\mbox{fb$^{-1}$}}
\def\ipb{\mbox{pb$^{-1}$}}
\def\pt{\ensuremath{p_{\mathrm{T}}}}
\def\jpsi{\ensuremath{J/\psi}}
\def\Ups{\ensuremath{\mit\Upsilon}}

\begin{document}

\begin{minipage}[t]{\textwidth}
  \begin{flushright}
    CERN-PH-EP-2011-219\\
    Submitted to Physical Review D
    \end{flushright}
\end{minipage}
\vspace*{0.5cm}

\title{Search for anomalous production of prompt like-sign muon pairs and constraints on physics beyond the Standard Model with the ATLAS detector}

\author{The ATLAS Collaboration}

\begin{abstract}
  An inclusive search for anomalous production of
  two prompt, isolated muons with the same electric charge is
  presented.  The search is performed in a data sample corresponding
  to 1.6 \ifb\ of integrated luminosity collected in 2011 at $\sqrt{s}=7$~TeV 
  with the ATLAS detector at the LHC. 
  Muon pairs are selected by requiring two isolated muons of the same electric charge 
  with $\pt>20$~GeV and $|\eta|<2.5$. 
  Minimal requirements are placed on the rest of the event activity.
  The distribution of the invariant mass of the muon pair $m(\mu\mu)$
  is found to agree well with the background expectation.
  Upper limits on the cross section for anomalous production of two muons with the same electric charge are placed 
  as a function of $m(\mu\mu)$ within a fiducial region defined by the event selection.
  The fiducial cross-section limit constrains the  like-sign top-quark pair-production 
  cross section to be below 3.7~pb at 95\% confidence level.
  The data are also analyzed to search for a narrow like-sign dimuon resonance as 
  predicted for e.g. doubly charged Higgs bosons ($\hpp$).
  Assuming pair production of $\hpp$ bosons and
  a branching ratio to muons of 100\% (33\%), this analysis excludes 
  masses below 355 (244) GeV and 251 (209) GeV for $\hpp$ bosons coupling 
  to left-handed and right-handed fermions, respectively.
\end{abstract}


\maketitle

\clearpage

\section{Introduction}
\label{sec:intro}

Events containing two high-$\pt$, prompt, like-sign leptons are rarely
produced in the Standard Model (SM), but occur with an enhanced rate in
several models of new physics.  For example, supersymmetry~\cite{theorySUSY},
universal extra dimensions~\cite{theoryUED}, 
left-right symmetric models~\cite{LRSM1,LRSM2,LRSM3,LRSM4}, 
Higgs triplet models~\cite{Hpp_triplet,triplet1,triplet2}, the little Higgs model~\cite{hpp_little},
fourth-family quarks~\cite{theory4th}, and flavor-changing neutral currents resulting in the 
production of like-sign top 
quarks~\cite{murayama,cao,Choudhury:2010cd,Cao:2011ew,Bhattacherjee:2011nr,berger,AguilarSaavedra:2011zy,ko1,ko2}
could all give rise to final states with two leptons of the same electric charge. 
Most of these models would result in an excess of like-sign dimuons over the background with 
no distinct kinematic features. However, doubly charged Higgs bosons ($\hpp$), 
predicted by some of those models, would be observed as a narrow resonance 
in the dimuon mass spectrum.

In the analysis described in this article, events containing like-sign muon pairs are selected and their invariant mass 
distribution is compared to the SM prediction. Both muons are required to have transverse momentum
$\pt>20$~GeV and pseudorapidity~\footnote{ATLAS uses a right-handed coordinate system with its origin 
at the nominal interaction point (IP) in the center of the detector and the $z$-axis 
along the beam pipe. The $x$-axis points from the IP to the center of the LHC ring, and 
the $y$ axis points upward. Cylindrical coordinates $(r,\phi)$ are used in the transverse 
plane, $\phi$ being the azimuthal angle around the beam pipe. The pseudorapidity is defined 
in terms of the polar angle $\theta$ as $\eta=-\ln\tan(\theta/2)$.}
$|\eta|<2.5$, and they must be isolated from other activity in the event. 
Upper limits on the cross section of non-SM physics 
in a fiducial region corresponding to the experimental requirements are derived as a function 
of the dimuon invariant mass. Results are presented inclusively for $\mu^\pm\mu^\pm$ 
production and separately for $\mu^+\mu^+$ and $\mu^-\mu^-$ final states. The $\mu^+\mu^+$ result is 
further used to constrain like-sign top-quark pair production.  The data are also used 
to search for a narrow dimuon resonance with a width much smaller than the detector resolution of 
$\sim3\%$.  An example of a particle that may result in a narrow mass peak is a short-lived
$\hpp$ boson, predicted by a number of the models for new physics mentioned above.
Constraints on the $\hpp$ mass as a function of its branching ratio to two muons 
are presented.

The ATLAS Collaboration has previously reported an inclusive search for new physics in
the like-sign dilepton final state in a data sample corresponding to
an integrated luminosity of 34~$\ipb$~\cite{2010SS}.  No significant deviation from SM
expectations was observed, and fiducial cross-section limits as well
as limits on several specific models of physics beyond the SM were
derived. The CDF Collaboration has performed similar inclusive searches~\cite{cdf,cdf2} without
observing any evidence for new physics.  Like-sign top-quark pair production has previously been searched for by the CDF ~\cite{cdfsstop}
and the CMS Collaborations ~\cite{cmssstop}. The upper limit on the cross section set by the 
CMS Collaboration in $pp$ collisions at $\sqrt{s}=7$~TeV is 17~pb.  Direct limits on $\hpp$ bosons have previously 
been set at hadron colliders by the CDF~\cite{hpp_cdf, cdf2} and D0~\cite{hpp_d0,hpp_d0_2}
Collaborations. The most stringent limits to date for $\hpp$ 
bosons decaying to dimuons with a branching ratio of 100\% exclude masses below 
205-245~GeV depending on the couplings~\cite{cdf2}.

This article is organized as follows. A brief description of the ATLAS detector 
is given in Section~\ref{sec:atlas}. Section~\ref{sec:datamc} presents the 
data and simulation samples used. The event selection is described in Section~\ref{sec:evsel}. 
The backgrounds are
discussed in Section~\ref{sec:backgrounds}, and Section~\ref{sec:systematics}
summarizes the systematic uncertainties. The data are compared to the background 
estimate in Section~\ref{sec:results}. The interpretation of the data as a  
cross-section upper limit within the fiducial region, for four ranges of dimuon invariant mass,
and its implication on like-sign top-quark pair production are reported in Sections~\ref{sec:fidlimit}
and \ref{sec:sstop}, respectively. 
The narrow resonance search and its interpretation in terms of $\hpp$ boson production
is presented in Section~\ref{sec:hpp}.
Finally, Section~\ref{sec:conclusions} summarizes the conclusions.

\section{The ATLAS detector}
\label{sec:atlas}

The ATLAS detector~\cite{atlas} consists of an inner tracking system, 
calorimeters, and a muon spectrometer.
The inner detector, directly surrounding the interaction point, is composed of a silicon pixel 
detector, a silicon strip detector, and a transition radiation tracker, all
embedded in a 2~T axial magnetic field. It covers the pseudorapidity 
range $|\eta|<2.5$ and is enclosed by a calorimeter system containing electromagnetic 
and hadronic sections. The calorimeter system is surrounded by a large muon 
spectrometer built with three air-core toroids.
This spectrometer is equipped with precision chambers (composed of monitored drift tubes and cathode strip chambers) to provide 
precise position measurements in the bending plane in the range $|\eta|< 2.7$.
In addition, resistive plate chambers and thin gap chambers with 
a fast response time are used primarily to trigger muons in the 
rapidity ranges $|\eta|\le 1.05$ and $1.05 <|\eta|< 2.4$, respectively.
Momentum measurements in the muon spectrometer are based on track segments formed in at least two 
of the three precision chambers. The resistive plate chambers and thin gap chambers 
provide position measurements in the non-bending plane which is used to improve the 
pattern recognition and the track reconstruction.

The ATLAS detector has a three-level trigger system~\cite{triggerpaper}
which reduces the event rate to approximately 200 Hz before
data transfer to mass storage. 
The Level-1 muon trigger searches for hit coincidences between different 
muon trigger detector layers inside programmed geometrical windows that define the muon transverse 
momentum and provide a rough estimate of its position. It selects muons in the rapidity range $|\eta|<2.4$.  
The Level-1 trigger is followed by a high-level, software-based trigger selection 
which is similar to that of the offline reconstruction.

\section{Data sample and Monte Carlo simulation}
\label{sec:datamc}

This analysis is carried out using a data sample corresponding to an integrated 
luminosity of 1.6 \ifb\, recorded between
March and July of 2011 at a center-of-mass energy of 7~TeV. The data are selected using single-muon triggers 
with a $\pt$ threshold of 10 GeV at Level-1.
At the high-level trigger, a muon with $\pt>18$~GeV is required.
In this dataset, the average number of interactions per beam crossing is about six. 

Monte Carlo (MC) simulation is used to estimate some of the background contributions 
and to determine the selection efficiency and acceptance for possible new physics signals. The dominant 
SM processes that contribute to prompt like-sign dimuon production
are $WZ$, $ZZ$, $W^\pm W^\pm$, and $t\bar{t}W$. These are all estimated using MC simulation.  
For processes with a $Z$ boson, the contribution from $\gamma^*$ is also 
simulated for $m(\ell\ell)>20$~GeV. $WZ$ and $ZZ$ events are generated using \textsc{Herwig}~\cite{herwig}, and
$W^\pm W^\pm$ and $t\bar{t}W$ production is generated with {\sc MadGraph}~\cite{madgraph} 
for the matrix element and {\sc Pythia}~\cite{pythia} for the parton shower and fragmentation.  

The normalization of the $WZ$ and $ZZ$ MC samples is based on cross
sections determined at next-to-leading-order (NLO) using {\sc
  MCFM}~\cite{mcfm}. The NLO cross sections times branching ratios for
$W^{\pm}Z \rightarrow \ell^{\pm}\nu\ell^{\pm}\ell^{\mp}$ and $ZZ \rightarrow
\ell^{\pm}\ell^{\mp}\ell^{\pm}\ell^{\mp}$, where $\ell^{\pm}$ is an electron, muon, or tau lepton, 
after requiring two charged
leptons with the same electric charge and with $\pt>20$~GeV and $|\eta|<2.5$, are $347$~fb 
and $54$~fb, respectively. The $K$-factors for WZ and ZZ production, defined as the ratios between the NLO and the
leading order (LO) cross sections, depend on the kinematic requirements
placed on the muons and the invariant mass of the like-sign muon
pair. Therefore, $K$-factors that depend on this invariant mass are
applied.

Opposite-sign dimuon events due to Drell-Yan, $t\bar{t}$, and $W^\pm W^\mp$
production constitute a background if the charge of 
one of the muons is misidentified. $W^\pm W^\mp$ production is generated using \textsc{Herwig}.
The Drell-Yan process is generated with \textsc{Alpgen}~\cite{alpgen}, whereas the $t\bar{t}$ 
background is modeled using \textsc{MC@NLO}~\cite{mcatnlo}.

In addition, a variety of new physics signals are simulated in order to study the efficiency and 
acceptance of the selection cuts. 

Like-sign top-quark pair production can occur in models 
with flavor-changing neutral currents, e.g. via a $t$-channel exchange of a $Z'$ boson with $utZ'$ coupling. 
Since the left-handed coupling is highly constrained by $B^{0}_d-\bar{B}^{0}_d$ 
mixing~\cite{bd}, only right-handed top quarks ($t_R$) are considered. 
Samples for this process are produced with the {\sc Protos}~\cite{Protos} generator, 
using $Z'$ mass values of 100, 150 and 200 GeV. An additional sample is generated, based on an effective four-fermion 
operator $uu \to tt$ corresponding to $Z'$ masses $\gg 1$~TeV~\cite{AguilarSaavedra:2011zy}. The parton shower and 
hadronization are performed with {\sc Pythia}. 

Pair production of doubly charged Higgs bosons ($pp\to \hpp H^{\mp\mp}$) via a virtual $Z/\gamma^*$ exchange
is generated using {\sc Pythia} for $\hpp$ mass values between 100 and 
400~GeV~\cite{theory_hpphmm}.

Production of a right-handed $W$ boson ($W_R$) 
decaying to a charged lepton and a Majorana neutrino ($N_R$) ~\cite{theory_LRSM,theoryMNeu},
and pair production of heavy down-type fourth generation quarks ($d_4$) decaying to $tW$
are generated using {\sc Pythia}.

Parton distribution functions taken from CTEQ6L1~\cite{cteq6l1} are used for the LO MC generators, while for 
the $t\bar{t}$ \textsc{MC@NLO} sample CTEQ6.6~\cite{cteq66} parton distribution functions are used.

The detector response to the generated events is simulated with the ATLAS simulation framework~\cite{atlassim} 
using \textsc{Geant4}~\cite{g4}, and the events are reconstructed with the same 
software used to process the data.  The simulated response is corrected for the small 
differences in efficiencies, momentum scales, and momentum resolutions between data and simulation.

\section{Event selection}
\label{sec:evsel}

Events are selected with an inclusive single-muon trigger with a $\pt$ threshold of 18~GeV 
as described in Section~\ref{sec:datamc}.
They must further contain at least two muons of the same electric charge 
with $\pt>20$~GeV and $|\eta|<2.5$.
The efficiency of the trigger selection for muon pairs in $Z \to \mu^+\mu^-$ events passing the event selection used here is $~97\%$.  
Any combination of two muons is considered, allowing more than one
muon pair per event to be included. The invariant mass of the two muons,
$m(\mu\mu)$, is required to be larger than 15~GeV to exclude the 
low-mass hadronic resonances such as the $\jpsi$ and $\Ups$ mesons. 
All events used in this analysis are required to have a 
primary vertex determined with at least five tracks with $\pt>0.4$~GeV. If more 
than one interaction vertex is found, the vertex with the highest 
$\sum_{i=1}^{N} p_{{\rm T},i}^2$, where $N$ is the number of tracks associated to the vertex,
is defined as the primary vertex.

Muons selected for this analysis are formed from tracks reconstructed in the inner detector combined 
with tracks reconstructed in the muon spectrometer~\cite{staco}. The independent charge measurements from 
these two detectors are required to agree to reduce the charge mismeasurement rate. In addition, the transverse and longitudinal impact 
parameters with respect to the primary event vertex must be small, $|d_0|<0.2$~mm and 
$|z_0\sin{\theta}|<5.0$~mm, and the transverse impact parameter 
significance, $\dzsig$, is required to be less than 3.0. The muon isolation ($\pt^{{\rm cone}40}$) is defined 
as the scalar sum of the transverse momenta of all tracks with $\pt>0.5$~GeV 
within a cone around the muon axis of size $\Delta R = \sqrt{\Delta\phi^2+\Delta\eta^2} = 0.4$
that are within $|z_0|<1$~cm of the primary event vertex. 
Requirements of $\pt^{{\rm cone}40}<5$~GeV and $\pt^{{\rm cone}40}/\pt(\mu)<0.08$ are made. 

The above selection cuts are chosen to retain a high efficiency for prompt muons while rejecting 
a large fraction of non-prompt backgrounds. For muons from $Z$-boson decays, the efficiency of the impact parameter 
significance and the isolation cuts ranges from 87\% to 97\% depending on $\pt$, 
while for muons from $b$- and $c$-hadron decays, the efficiency is about 3.5\%. For muons arising 
from $\tau$ decays in $Z\to\tau\tau$ events, the efficiency is about 60\%.

\section{Background determination}
\label{sec:backgrounds}

The SM backgrounds for like-sign dimuon final states can be divided into background from production of
prompt like-sign dimuons, background caused by muons from hadronic
decays (non-prompt muons), and background from processes with two prompt
opposite-sign muons where the charge of one of the muons is mismeasured. 

The dominant SM processes with two prompt leptons of the same
electric charge in the final state are $W^{\pm}Z \rightarrow \ell^\pm\nu \ell^\pm\ell^\mp$, 
$ZZ \rightarrow \ell^\pm\ell^\mp\ell^\pm\ell^\mp$, $W^\pm W^\pm\to
\ell^\pm \nu \ell^\pm \nu$, and $t\bar{t}W \to \ell^\pm\ell^\pm+X$. 
Any other SM processes are found to be negligible. 
The contribution of these processes to the signal region is estimated from MC simulation
using the samples described in Section~\ref{sec:datamc}.
In these simulated samples, only muons that originate from a $\tau$ lepton, a $W$ boson, 
or a $Z$ boson are considered prompt. Muons originating from any other sources are discarded 
in order to avoid double-counting with the non-prompt muon background that is derived from data. 

Background from non-prompt muons may originate from several
different sources: semi-leptonic $b$- or $c$-hadron
decays, muons from pion or kaon decays in flight, and misidentified 
muons from hadronic showers in the calorimeter which reach the muon spectrometer and are incorrectly 
matched to a reconstructed inner detector track~\cite{incllep}.
The background from non-prompt muons is estimated from data using a matrix method~\cite{MatrixMethod}. 
This method requires knowledge of the probabilities for prompt and non-prompt muons to pass 
the isolation requirement. The probability for non-prompt muons to pass the isolation cut
is determined using muons with $\dzsig>5$ in dimuon or single muon samples. These are 
dominated by semi-leptonic $b$- and $c$- hadron decays.
The probability is found to be 5\% rather independently of $\pt$ and $\eta$. 
A systematic uncertainty is derived from a complementary sample where $\dzsig<3$ is required. In this sample, 
prompt muons from $W$ or $Z$ decays are suppressed by requiring there to be exactly one muon in the event, 
the transverse mass~\footnote{The transverse mass is defined as 
$\sqrt{2E_{\rm T}^{\rm miss}p_{\rm T}(\mu)(1-\cos\Delta\phi(\mu,E_{\rm T}^{\rm miss}))}$ where $E_{\rm T}^{\rm miss}$ 
is the missing transverse energy as explained
in Ref.~\cite{metpaper} and $\Delta\phi(\mu,E_{\rm T}^{\rm miss})$ is the difference between the azimuthal angles of the 
muon and $E_{\rm T}^{\rm miss}$.} 
of the muon and the missing transverse energy~\cite{metpaper} to be below $10$~GeV, 
and at least one jet with $\pt>20$~GeV to be present. 
The resulting systematic uncertainty on the probability for non-prompt muons to pass the isolation cut varies 
between 30\% and 100\% depending on $\pt$. 
The probability for prompt muons to pass the isolation cut as a function of $\pt$ and $\eta$ 
is derived from $Z \to \mu^+\mu^-$ MC events and is cross-checked with data.

Another source of background arises from opposite-sign muon pairs where the charge of one 
of the two muons is misidentified. This background source is negligible
in the relevant mass range as estimated from simulation.
The charge misidentification probability is also measured from $Z\to\mu\mu$ events in data 
by exploiting the independent charge measurements provided by the inner detector and the muon spectrometer. 
It is found to be consistent with zero in the relevant $\pt$ range.  Based on observing zero charge misidentified events in data, 
a 68\% upper limit is placed on this probability
as function of $\pt$, which ranges up to 10\% at $\pt(\mu)=400$~GeV. 
This upper limit is applied as a function of $\pt(\mu)$ to opposite-sign prompt 
muon pairs in the Drell-Yan, $W^+ W^-$, and $t\bar{t}$ 
MC samples to determine the systematic uncertainty on this background source.

The background estimate is cross-checked in a variety of samples complementary to the signal region.
These include like-sign muon pairs where at least one muon fails the $\dzsig$ cut, 
like-sign muon pairs where both muons fail the isolation requirement 
used in the analysis but pass a looser isolation requirement, 
like-sign and opposite-sign muon pairs where both muons fail the isolation requirement 
used in the analysis but pass a looser isolation requirement and at least one muon fails the $\dzsig$ cut, 
and opposite-sign muon pairs where both muons pass the final analysis requirements. 
For all control regions, the data are found to agree with the background prediction 
within the systematic uncertainties, both in overall event yield and in the shape of the dimuon mass distribution.

\section{Systematic uncertainties}
\label{sec:systematics}

Uncertainties on the event selection efficiencies and the luminosity affect the predicted 
yield of signal events as well as those backgrounds that are estimated purely from MC simulation, 
i.e. $WZ$, $ZZ$, $W^\pm W^\pm$, and $t\bar{t}W$ production. The uncertainty on the muon reconstruction 
efficiency is $\pm 1\%$~\cite{wwxs}. In addition, the efficiency of the requirements on 
impact parameter and isolation is observed to be 3\% lower in data than in simulation at the lowest $\pt$ values 
while for $\pt>30$~GeV data and simulation agree typically within $\pm 1\%$.
The resulting uncertainty on the muon pair selection efficiency due to the muon identification efficiency
is $^{+1.0}_{-1.8}\%$.
The uncertainty on the muon trigger efficiency of $<1\%$~\cite{wwxs} 
results in an uncertainty on the selection efficiency of $\pm 0.3\%$.
The uncertainty in the muon momentum scale~\cite{momscale} results in an
uncertainty on the dimuon pair selection efficiency of $\pm 0.9\%$ due to the 
migrations across the $\pt$ and $m(\mu\mu)$ cut thresholds.
In addition, the integrated luminosity measurement has an uncertainty of $\pm 3.7$\%~\cite{lumi,lumipaper}.

The uncertainty in the production cross sections of the SM processes 
affect the predicted yield of the prompt muon background. 
The $WZ$ and $ZZ$ cross-section uncertainties due to higher-order 
corrections are estimated to be $\pm 10\%$ by varying the renormalization and 
factorization scales by a factor of two. 
For $t\bar{t}W$ production, the higher-order corrections are estimated to be similar to those 
for $t\bar{t}Z$, which are calculated in Ref.~\cite{nlottw}, 
and the cross section is taken to be a factor of $1.30\pm 0.65$ higher than the LO cross 
section~\cite{petriello}. The full higher-order corrections for $W^{\pm}W^{\pm}$ production 
have not yet been calculated. However, for parts of the process, the NLO QCD corrections have been shown 
to be small~\cite{nlowpwp}. Here, the LO cross section is used and an uncertainty of $\pm 50\%$ is assumed.

Uncertainties on the parton distribution functions affect both the acceptance and the normalization of 
the prompt muon backgrounds and the new physics models
constrained in this paper. This uncertainty is evaluated using the eigenvectors
provided by the {\tt MSTW2008lo68cl} set~\cite{mstw2008} of parton distribution functions
using the prescription given in Ref.~\cite{pdferr_new} and adding in 
quadrature the difference between the central cross-section value obtained using
this set and that obtained with the {\tt CTEQ6L1}~\cite{cteq6l1} parton distribution functions.
For the diboson background, the resulting uncertainty on the cross section is $\pm 7\%$. 
The uncertainty on the acceptance due to this source is typically $\pm 2\%$.

The uncertainty on the number of muon pairs from non-prompt muon backgrounds has systematic and 
statistical components which are added in quadrature
to give the total uncertainty on this background source. The systematic component is derived from 
the uncertainty on the measurement of the fraction of non-prompt muons passing the isolation cuts which ranges from $\pm30\%$ for $m(\mu\mu)>15$~GeV to $\pm80\%$ for $m(\mu\mu)>300$~GeV (see Section~\ref{sec:backgrounds}). 
The statistical component arises from the limited number of non-isolated muons used in the 
matrix method: this is $\pm3\%$ for $m(\mu\mu)>15$~GeV and $\pm45\%$ for $m(\mu\mu)>300$~GeV.
The background due to charge misidentification has an uncertainty of $+ 2.7$ events for the full 
sample and $+0.6$ events in the highest mass region.

Any statistical uncertainties due to limited size of the background and signal 
MC samples are also considered.
 
Systematic uncertainties on different processes from the same origin
are assumed to be 100\% correlated.

\section{Comparison of the data to the background expectation}
\label{sec:results}

The invariant mass distributions observed in the data are compared to the predicted background 
for $\mu^\pm\mu^\pm$, $\mu^+\mu^+$, and $\mu^-\mu^-$ production in Figure~\ref{fig:mass}. 

\begin{figure}[htbp]
\begin{center}
\subfigure[]{\includegraphics[width=0.41\textwidth]{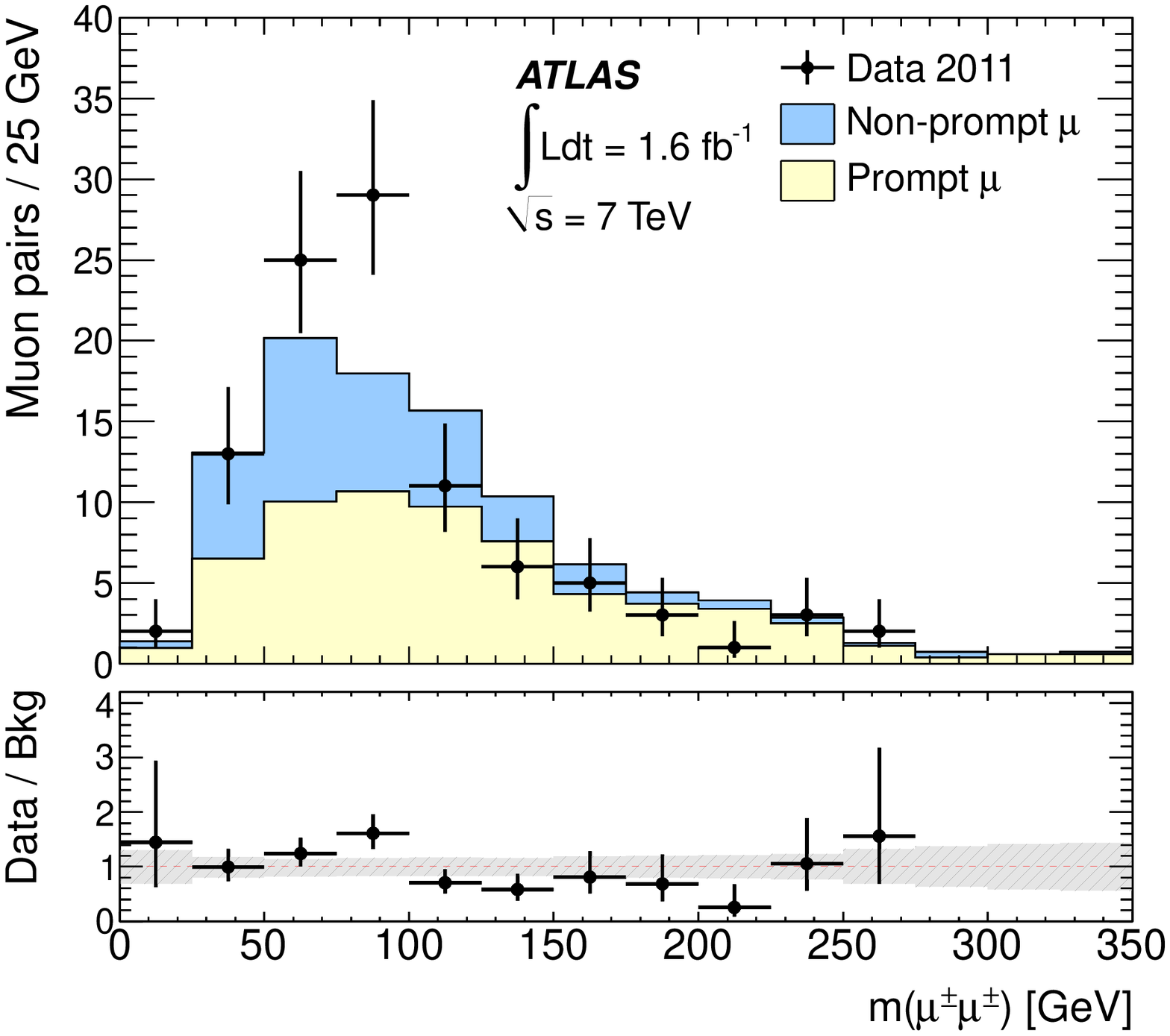}}
\subfigure[]{\includegraphics[width=0.41\textwidth]{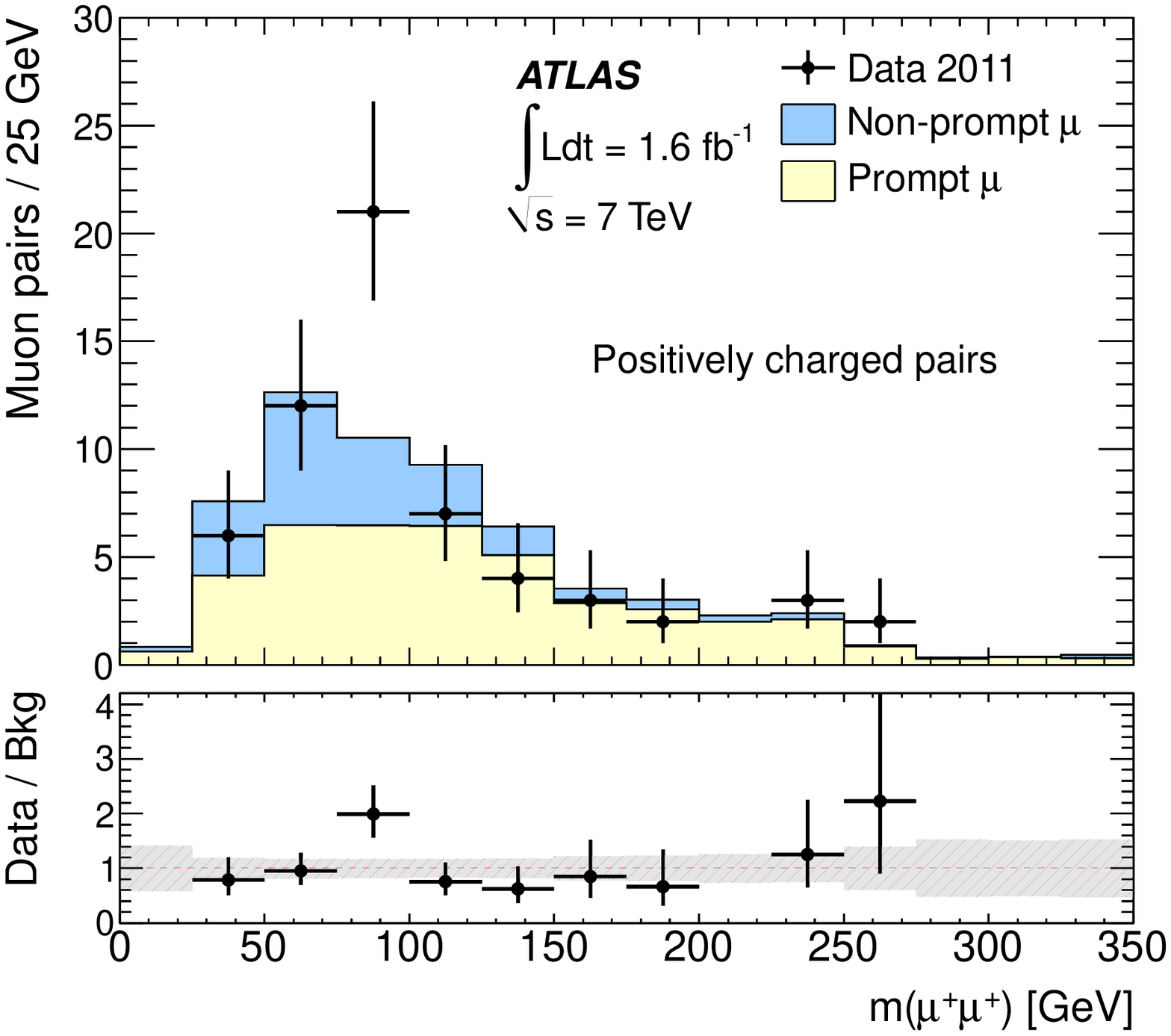}}
\subfigure[]{\includegraphics[width=0.41\textwidth]{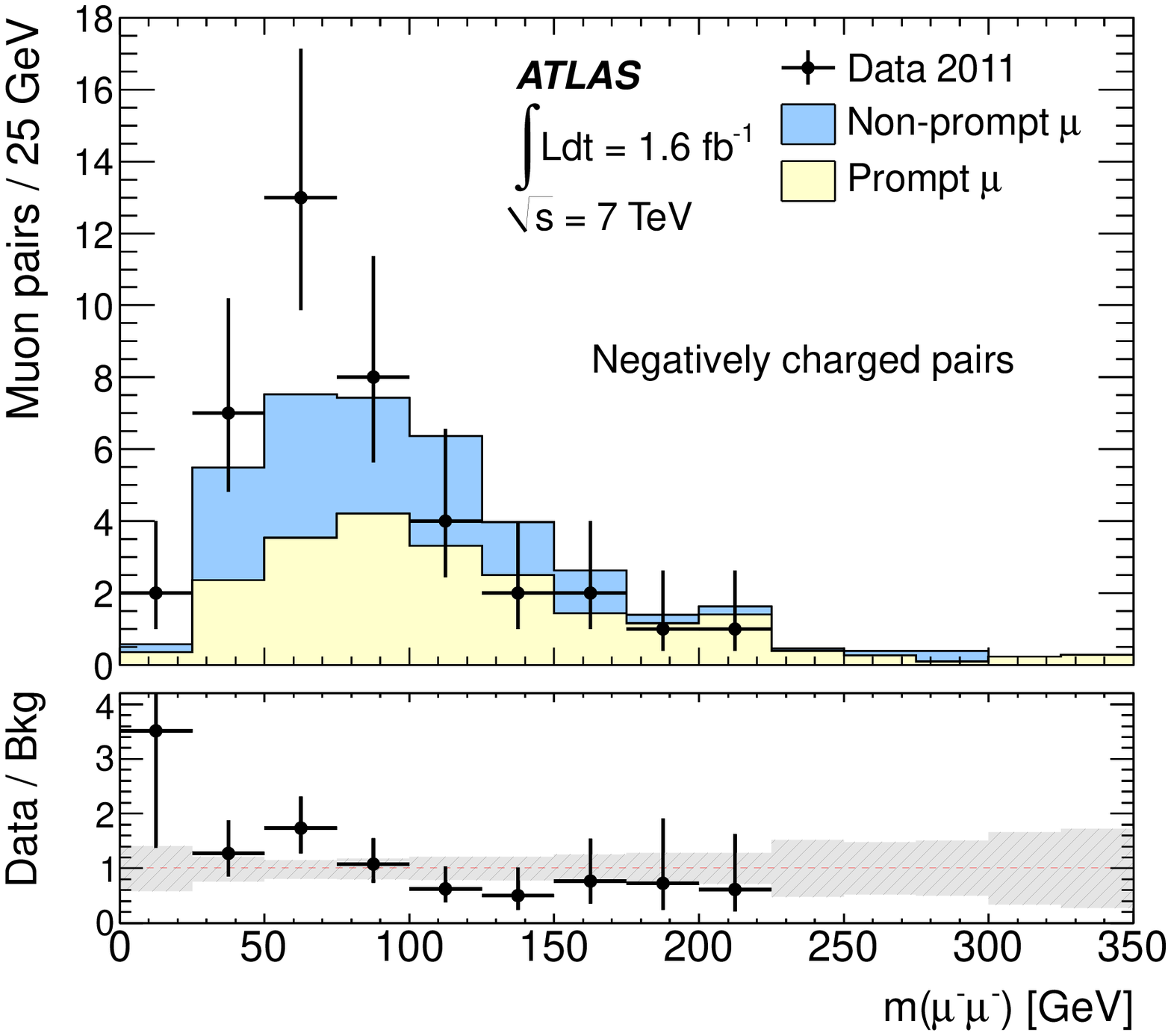}}
\caption{\label{fig:mass}  
Distribution of the dimuon invariant mass for a) $\mu^\pm\mu^\pm$ pairs, b) $\mu^+\mu^+$ pairs, and c) $\mu^-\mu^-$ pairs.
The data are compared to the stacked background estimates. 
The ratio between the data and the predicted background 
is also shown, where the shaded region is the total systematic uncertainty on the 
background prediction.}
\end{center}
\end{figure}

Table~\ref{tab:databg} summarizes the number of observed and expected muon pairs 
for $\mu^\pm\mu^\pm$, $\mu^+\mu^+$, and $\mu^-\mu^-$ 
production for four cuts on the dimuon invariant mass. The data 
agree with the background within the systematic uncertainties and no excess is observed. 
The number of data events in high-mass bins is lower than the background expectation, but in all mass
bins the probability that the background gives a fluctuation as low or lower than observed in the data is found to 
be greater than $5\%$. In all mass bins, prompt muons from diboson production are the dominant background
but non-prompt muons also contribute significantly: about 40\% at low mass and 10\% at high mass.

\begin{table*}[htbp]
\caption{Expected and observed numbers of pairs of isolated like-sign muons 
for various cuts on the dimuon invariant mass, $m(\mu\mu)$. 
The uncertainties shown are the quadratic sum of the statistical and systematic uncertainties.
The prompt muon background contribution includes the $WZ$, $ZZ$, $W^{\pm}W^{\pm}$, and $t\bar{t}W$ processes.}
\begin{center}
\begin{tabular}{l|c|c|c|c}
\hline
Sample & \multicolumn{4}{|c}{Number of muon pairs with  $m(\mu^{\pm}\mu^{\pm})$} \\
 & $>15$~GeV & $>100$~GeV & $>200$~GeV & $>300$~GeV \\
\hline \hline
prompt muons         & $63.1 \pm 7.8$      & $34.9 \pm 4.5$ & $9.6 \pm 1.6$ & $2.24 \pm 0.54$ \\[+0.05in]
non-prompt muons & $37.5^{+10.3}_{-12.4}$ & $13.0 \pm 4.5$ & $1.8 \pm 0.7$ & $0.31 \pm 0.18$ \\[+0.05in]
charge flip               & $0^{+2.7}_{-0.0}$    & $0^{+0.9}_{-0.0}$  & $0^{+0.7}_{-0.0}$ & $0^{+0.61}_{-0.00}$ \\[+0.05in]
\hline
total              & $100.6^{+13.2}_{-14.7}$ & $48.0 \pm 6.4$ & $11.4^{+1.8}_{-1.7}$ & $2.56^{+0.83}_{-0.57}$ \\[+0.05in]\hline \hline
data             & 101 & 32 & 7 & 1 \\ \hline
\end{tabular}
\\ \vspace*{0.8cm} 
\begin{tabular}{l|c|c|c|c}
\hline
Sample & \multicolumn{4}{|c}{Number of muon pairs with  $m(\mu^+\mu^+)$} \\
 & $>15$~GeV & $>100$~GeV & $>200$~GeV & $>300$~GeV \\
\hline \hline
prompt muons     & $41.2 \pm 5.3$ & $23.5 \pm 3.2$ & $6.6 \pm 1.2$ & $1.33 \pm 0.40$ \\[+0.05in]
non-prompt muons & $20.2^{+5.9}_{-6.9}$ & $6.3 \pm 2.2$ & $1.0 \pm 0.4$ & $0.24 \pm 0.15$ \\[+0.05in]
charge flip      & $0^{+1.3}_{-0.0}$ & $0^{+0.5}_{-0.0}$ & $0^{+0.3}_{-0.0}$ & $0^{+0.30}_{-0.00}$ \\[+0.05in]
\hline
total              & $61.4^{+8.0}_{-8.7}$ & $29.8 \pm 3.9$ & $7.5 \pm 1.3$ & $1.57^{+0.52}_{-0.42}$ \\[+0.05in]\hline \hline
data             & 61 & 22 & 6 & 1 \\
\hline
\end{tabular}
\\ \vspace*{0.8cm} 
\begin{tabular}{l|c|c|c|c}
\hline
Sample & \multicolumn{4}{|c}{Number of muon pairs with  $m(\mu^-\mu^-)$} \\
 & $>15$~GeV & $>100$~GeV & $>200$~GeV & $>300$~GeV \\
\hline \hline
prompt muons     & $21.9 \pm 3.0$ & $11.4 \pm 1.8$ & $3.04 \pm 0.67$ & $0.91 \pm 0.32$ \\[+0.05in]
non-prompt muons & $17.4^{+4.7}_{-5.8}$ & $6.8\pm 2.4$ & $0.83 \pm 0.38$ &  $0.07^{+0.08}_{-0.07}$ \\[+0.05in]
charge flip      & $0^{+1.3}_{-0.0}$ & $0^{+0.5}_{-0.0}$ & $0^{+0.34}_{-0.0}$ & $0^{+0.30}_{-0.00}$ \\[+0.05in]
\hline
total              & $39.3^{+5.8}_{-6.5}$ & $18.2 \pm 3.0$ & $3.87^{+0.84}_{-0.77}$ & $0.98^{+0.45}_{-0.33}$ \\[+0.05in]\hline \hline
data             & 40 & 10 & 1 & 0 \\ \hline
\end{tabular}
\end{center}
\label{tab:databg}
\end{table*}

\section{Upper limits on the cross section for prompt like-sign dimuon production}
\label{sec:fidlimit}

A 95\% confidence level (C.L.) upper limit on the number of like-sign muon pairs due to anomalous production, 
$N_{95}(\mu\mu)$, is obtained using a Bayesian approach with a flat prior for the number of events from new physics, 
integrating over Gaussian priors for the systematic uncertainties~\cite{mclimit,genlimit}.  All systematic uncertainties 
discussed above are included, and correlations between their effects on signal and background processes are taken into account.

The upper limit on the number of anomalously produced muon pairs, $N_{95}(\mu\mu)$, ranges from 
41 pairs for $m(\mu\mu)>15$~GeV to 3.8 pairs for $m(\mu\mu)>300$~GeV at 95\% C.L.
The limit on the number of muon pairs is translated  to a 95\% C.L. limit on the cross section
measured in the phase space region defined by the fiducial cuts as
\begin{equation}
\sigma^{fid}_{95}(\mu\mu) = \frac{N_{95}(\mu\mu)}{\varepsilon_{fid}\int{\cal L} {\rm d}t},
\end{equation}
where $\int{\cal L} {\rm d}t$ is the integrated luminosity of $1.61 \pm 0.06$~fb$^{-1}$. The efficiency
of the experimental cuts with respect to the fiducial region, $\varepsilon_{fid}$, depends on the model of new physics.
The fiducial cuts used to define the efficiency are closely matched to those imposed at reconstruction 
level: both muons must have $\pt>20$~GeV, $|\eta|<2.5$, and be separated by $\Delta R>0.4$ from any jet or 
prompt muon or electron with $\pt>20$~GeV.

A variety of models is considered for the determination of $\varepsilon_{fid}$, and the lowest 
efficiency value obtained among all the models is used. The models considered are like-sign top-quark 
pair production via an effective four-fermion coupling, Majorana neutrino ($N_R$) production from the decay 
of a $W_R$ boson, pair production of fourth generation quarks decaying via top quarks, and doubly charged Higgs 
boson production. A variety of mass values for those models is considered: $800\leq m(W_R)\leq 1500$~GeV 
and $100\leq m(N_R)\leq 1300$~GeV, $300\leq m(d_4)\leq 500$~GeV, and $100 \leq m(\hpp) \leq 300$~GeV.  
The efficiency values obtained from any of these samples with respect to the fiducial cuts vary for different models 
and mass bins due primarily to the $\pt$-dependence of the isolation efficiency.  Like-sign top-quark pair production 
results in the lowest fiducial efficiency of $43.9^{+1.9}_{-2.4}$\% for $m(\mu^\pm\mu^\pm)>300$~GeV, while a model 
with $W_R$ boson of 800~GeV decaying to a 500~GeV Majorana neutrino gives the highest
value of $72.5^{+1.6}_{-2.2}$\%.  For pair production of 100~GeV $\hpp$ bosons, the fiducial efficiency is 
$69.8^{+1.5}_{-2.0}$\% for $m(\mu^\pm\mu^\pm)>15$~GeV.
The efficiency uncertainties include all sources discussed in Section~\ref{sec:systematics}.
To derive the cross-section limits, the lowest efficiency value of $43.9^{+1.9}_{-2.4}$\% is used in all mass bins. 
The resulting limits are given in Table~\ref{tab:fidlim} for the four mass ranges and separately for 
$\mu^\pm\mu^\pm$, $\mu^+\mu^+$, and $\mu^-\mu^-$ production.

\begin{table}
\caption{Expected and observed 95\% C.L. upper limit on the cross section, $\sigma_{95}^{fid}$, 
for new physics in bins of dimuon mass for like-sign muon pairs with $\pt(\mu)>20$~GeV, $|\eta(\mu)|<2.5$, and 
$\Delta R>0.4$ between the muon and any jet, prompt electron or prompt muon with $p_T>20$~GeV.}
\begin{center}
\begin{tabular}{c| c|c}
\hline
\multirow{2}{*}{Mass range [GeV]} & \multicolumn{2}{|c}{$\sigma_{95}^{fid}$ [fb]} \\
 & expected & observed \\
\hline\hline
\multicolumn{3}{c}{All muon pairs} \\[+0.05in]
\hline
$m(\mu^{\pm}\mu^{\pm})>15$ & $58^{+19}_{-17}$ & 58 \\[+0.05in]
$m(\mu^{\pm}\mu^{\pm})>100$ & $30^{+11}_{-9}$ & 16 \\[+0.05in]
$m(\mu^{\pm}\mu^{\pm})>200$ & $13.7^{+5.7}_{-4.4}$ &  8.4 \\[+0.05in]
$m(\mu^{\pm}\mu^{\pm})>300$ & $8.0^{+3.3}_{-2.6}$ &  5.3 \\[+0.05in]
\hline
\hline
\multicolumn{3}{c}{Positively charged muon pairs} \\[+0.05in]
\hline
$m(\mu^{+}\mu^{+})>15$ & $37^{+14}_{-11}$ & 37 \\[+0.05in]
$m(\mu^{+}\mu^{+})>100$ & $21.8^{+9.1}_{-6.9}$ & 14.1 \\[+0.05in]
$m(\mu^{+}\mu^{+})>200$ & $10.3^{+5.7}_{-2.2}$ & 9.1 \\[+0.05in]
$m(\mu^{+}\mu^{+})>300$ & $7.2^{+1.8}_{-2.9}$ & 5.6 \\[+0.05in]
\hline
\hline
\multicolumn{3}{c}{Negatively charged muon pairs} \\[+0.05in]
\hline
$m(\mu^{-}\mu^{-})>15$ & $29^{+11}_{-8}$ & 30 \\[+0.05in]
$m(\mu^{-}\mu^{-})>100$ & $17.0^{+6.5}_{-5.1}$ & 9.5 \\[+0.05in]
$m(\mu^{-}\mu^{-})>200$ & $8.7^{+3.1}_{-2.5}$ & 5.2 \\[+0.05in]
$m(\mu^{-}\mu^{-})>300$ & $5.9^{+1.8}_{-1.6}$ & 4.3\\[+0.05in]
\hline
\end{tabular}
\end{center}
\label{tab:fidlim}
\end{table}

\section{Limits on like-sign top-quark pair production}
\label{sec:sstop}

Like-sign top-quark pair production can occur if e.g. a flavor-changing $Z'$ boson 
that couples to $u$ and $t$ quarks is exchanged in the $t$-channel.
The fiducial cross-section limits presented above are used to constrain this model.

In order to assess the impact on any physics model, the acceptance
of the fiducial cuts with respect to the full phase space, $A_{fid}$, needs to be determined. The cross-section 
limit for that model is then given by
\begin{equation}
\sigma_{95} = \frac{\sigma_{95}^{fid}(\mu\mu)}{A_{fid}}.
\end{equation}

For the model of like-sign top-quark production, only $\mu^+\mu^+$ pairs are considered
since the $\mu^-\mu^-$ process contributes less than 3\%
at the LHC due to the much smaller $\bar{u}$-quark density 
compared to the $u$-quark density in the proton. 
The fiducial acceptance for the production of right-handed like-sign top quarks, $A_{fid}(t_Rt_R)$, is determined for each mass cut 
and for four $Z'$ mass values.  For $m(Z')=100$~GeV ($m(Z')\gg 1$~TeV), $A_{fid}$ ranges from 
$0.69\%$ ($0.62\%$) for $m(\mu^+\mu^+)>15$~GeV to $0.12\%$ ($0.29\%$) for $m(\mu^+\mu^+)>300$~GeV.
This acceptance is defined with respect to inclusive decays of the $W$ bosons, so the small values are primarily caused 
by the low $W\to\mu\nu$ branching ratio.
The relative uncertainty on the acceptance is typically 2-3\% and accounts for
both the statistical uncertainty and the uncertainty due to the parton distribution functions as discussed 
in Section~\ref{sec:systematics}.  

The mass range that gives the best expected limits is $m(\mu^+\mu^+)>200$~GeV for all $m(Z')$.
The results are listed in Table~\ref{tab:ttlim} for four $Z'$ masses.
The upper limits on the $t_Rt_R$ production cross section range 
from 2.2 to 3.7~pb depending on $m(Z')$.

\begin{table}[htbp]
\caption{Upper limit at 95\% C.L. on the $t_Rt_R$ production cross section, $\sigma_{95}(t_Rt_R)$, for four $Z'$ mass 
values based on the $\mu^+\mu^+$ search with $m(\mu^+\mu^+)>200$~GeV.}
\begin{center}
\begin{tabular}{c|c|c}
\hline
\multirow{2}{*}{$m(Z')$} & \multicolumn{2}{|c}{$\sigma_{95}(t_Rt_R)$ [pb]} \\
 & expected & observed \\
\hline
100~GeV & $4.2^{+2.3}_{-0.9}$ & 3.7 \\[+0.05in]
150~GeV & $3.3^{+1.9}_{-0.7}$ & 3.0 \\[+0.05in]
200~GeV & $2.9^{+1.6}_{-0.6}$ & 2.6 \\[+0.05in]
$\gg 1$~TeV & $2.5^{+1.4}_{-0.5}$ & 2.2 \\[+0.05in]
\hline 
\end{tabular}
\end{center}
\label{tab:ttlim}
\end{table}

\section{Constraints on doubly charged Higgs bosons}
\label{sec:hpp}

The data are used to constrain the production of a narrow resonance decaying to two muons, using as reference model
the production of $\hpp$ bosons. In Section~\ref{sec:hppprod} the model considered 
for $\hpp$ production is described and the results are presented in Section~\ref{sec:hppresults}.

\subsection{$\hpp$ boson production}
\label{sec:hppprod}
The production process of doubly charged Higgs bosons considered here is pair production via the 
exchange of a virtual $Z/\gamma^*$~\cite{hpp_LR}. Other production mechanisms 
may contribute in addition but they depend on other model parameters such as the masses of the neutral and 
singly charged Higgs bosons and are therefore not included. Only $\hpp$ bosons decaying to muons with 
coupling values between $10^{-5}$ and $0.5$ 
are considered to ensure a short lifetime ($c\tau<10$~$\mu$m) 
and that the relative natural width, $\Gamma/M$, is less than $1\%$.  
Doubly charged Higgs bosons couple to Higgs and electroweak gauge bosons and either
left-handed or right-handed charged leptons, and are denoted $H^{\pm\pm}_L$ or $H^{\pm\pm}_R$, respectively. 
While $H^{\pm\pm}_L$ couple both to the $Z$ boson and to photons, $H^{\pm\pm}_R$ bosons only couple to photons,
i.e. coupling to any hypothetical right-handed gauge bosons is neglected,
resulting in a 2.5 times smaller pair-production cross section for the latter.

Next-to-leading-order calculations of the $\hpp$ pair-production cross section via the Drell-Yan process are used~\cite{spira}.
Higher-order QCD corrections beyond the next-to-leading-order accuracy are expected to increase the cross section by 
about 5\% but are neglected here.
The uncertainty on the cross section is $\pm10\%$ due to scale dependence in the NLO calculation, parton distribution 
function uncertainties,  and neglected electroweak corrections~\cite{spira_personal}.

\subsection{Constraints on $\hpp$ bosons}
\label{sec:hppresults}
The data are used to derive an upper limit on $\hpp$ pair production via the Drell-Yan process. 
For this purpose, counting experiments are performed in steps of 10 (20) GeV for 
$m(\mu\mu)<200$~GeV ($m(\mu\mu)\ge 200$~GeV) in a mass window of size $\pm10\%$ of the central mass, corresponding to about three times the experimental mass resolution.

The product of the acceptance and efficiency to detect a single $\hpp$ boson
is evaluated based on simulated samples.
It is $46\%$ at $m(\hpp)= 100$~GeV and increases to $57\%$ at 300~GeV. 
Uncertainties on the acceptance arise from the parton distribution functions, 
the interpolation between $\hpp$ mass values, and the limited MC statistics.
Adding these three uncertainties in quadrature, an overall acceptance 
uncertainty of $\pm3.6\%$ is obtained. The other systematic uncertainties are propagated as 
described in Section~\ref{sec:systematics}.

This analysis aims to constrain the pair production ($pp \to H^{++}H^{--}$) process. In the analysis, 
however, like-sign muon pairs are counted, and two muon pairs per event can contribute. 
The cross section for pair production of $\hpp$ bosons, $\sigma_{HH}$, is related to the number of reconstructed
dimuon pairs, $N(\mu^\pm\mu^\pm)$, by
\begin{equation}
\sigma_{HH}\times BR(\hpp \to \mu^\pm\mu^\pm)=\frac{N(\mu^\pm\mu^\pm)}{2\times A\times\epsilon\times {\cal L}{\rm d}t},
\end{equation}
where $A\times \epsilon$ is the acceptance times efficiency to detect a single $\mu^\pm\mu^\pm$ pair with invariant mass
within 10\% of the considered $\hpp$ mass value.  It was verified that for this process the efficiency for detecting a single $\mu^\pm\mu^\pm$ pair is not affected by the presence of a second pair in the event.

The cross-section limits are obtained using the same procedure as described in Section~\ref{sec:fidlimit}.
The expected and observed upper limits at 95\% C.L. on the cross section times 
the branching ratio, $\sigma(pp\to H^{++}H^{--}) \times BR(\hpp\to\mu^\pm\mu^\pm)$, are shown 
in Figure~\ref{fig:Hpplimit}. The observed upper limit is 11~fb at $m(\hpp)=100$~GeV and 
1.7~fb at $m(\hpp)=400$~GeV.
The median expected upper limits based on the background expectation together with the $\pm 1\sigma$ 
and $\pm 2\sigma$ uncertainty bands are also shown.
The results derived from data are consistent with the expectation over the full mass range.

The cross-section limit is compared to the prediction for the pair-production cross section 
of $\hpp_{L}$ and $\hpp_{R}$  bosons, assuming a branching ratio for 
the dimuon decay of 100\%. For this scenario, $\hpp_{L}$ bosons are excluded 
for $m(\hpp_L)<355$~GeV, while $\hpp_R$ bosons are excluded for $m(\hpp_R)<251$~GeV 
at 95\% C.L. for the central value of the theoretical prediction. 
The corresponding expected limits are 337~GeV and 264~GeV, respectively. 
Using a 10\% lower value for the theoretical prediction (corresponding to the $1\sigma$ 
uncertainty on the cross section), the data exclude $m(\hpp_L)<348$~GeV 
and $m(\hpp_R)<248$~GeV.
\begin{figure}[ht]
\begin{center}
\includegraphics[width=0.48\textwidth]{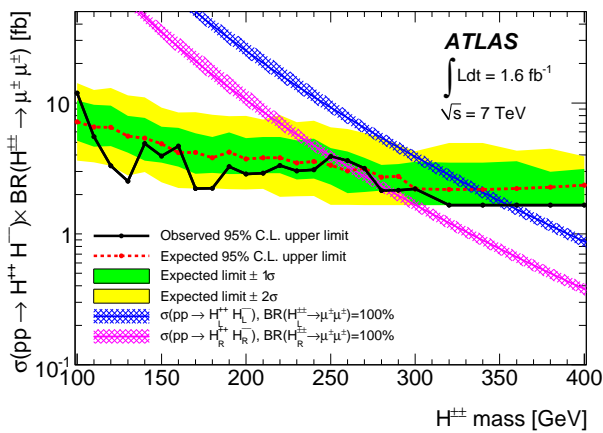}
\caption{\label{fig:Hpplimit} Upper limit at 95\% C.L. on the cross section times branching ratio 
for pair production of doubly charged Higgs bosons decaying to two muons.  Superimposed is the 
predicted cross section for $H^{++}_L H^{--}_L$ and $H^{++}_R H^{--}_R$ production 
assuming a branching ratio to muons of 100\%.  The bands on the predicted cross sections corresponds to 
the theoretical uncertainty of 10\%.}
\end{center}
\end{figure}

The observed and expected limits on the mass of doubly charged Higgs bosons are also determined 
as a function of the branching ratio to $\mu^\pm\mu^\pm$
assuming the central value of the theoretical cross-section prediction. 
This is shown in Figure~\ref{fig:limit_vsBR} for $\hpp_L$ and $\hpp_R$ bosons, respectively.  
For example, assuming a branching ratio of 33\% to muons, the respective lower mass limits are 
244~GeV for $\hpp_L$ and 209~GeV for $\hpp_R$ bosons.
\begin{figure*}[htbp]
\begin{center}
\subfigure[]{\includegraphics[width=0.48\textwidth]{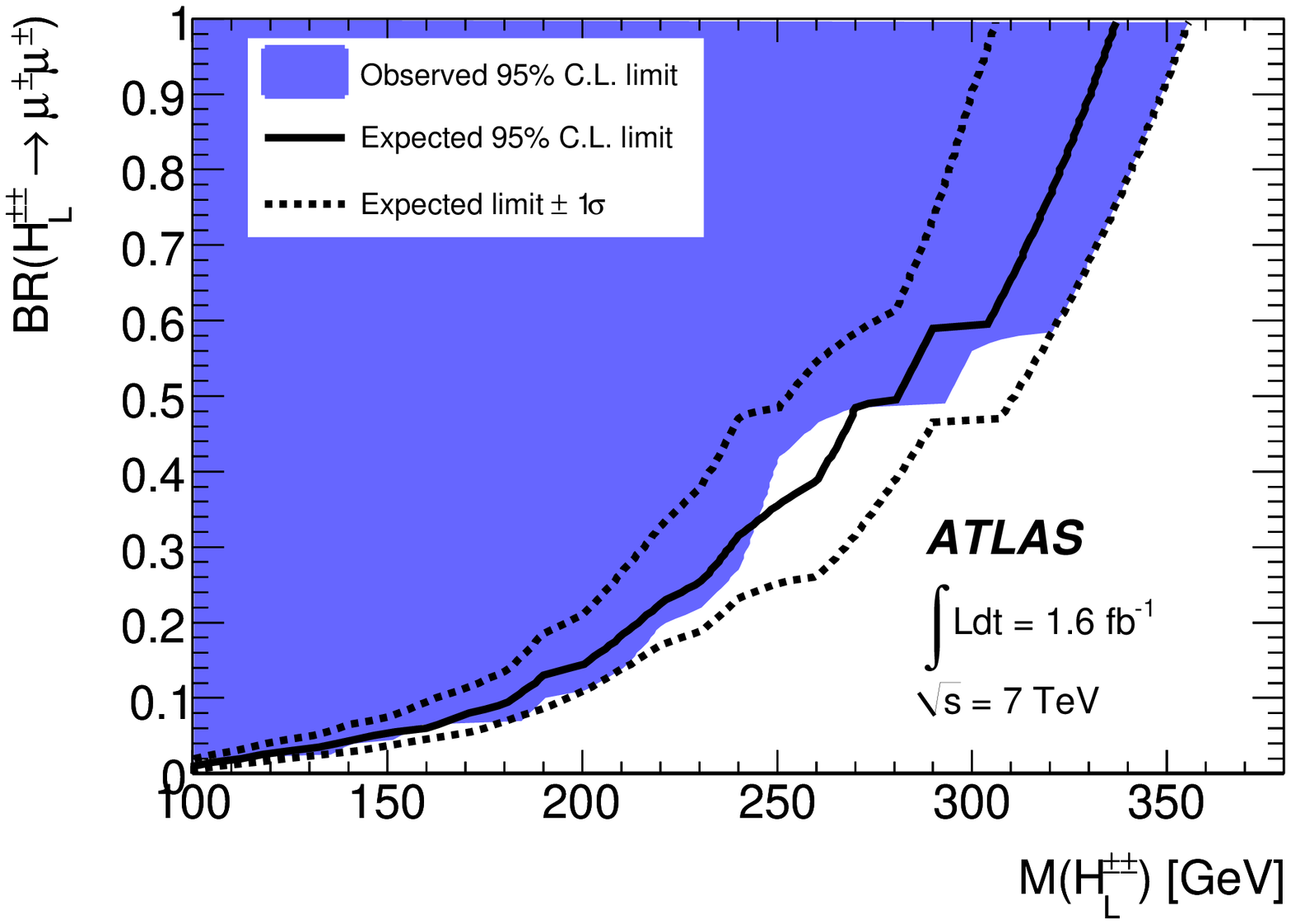}}
\subfigure[]{\includegraphics[width=0.48\textwidth]{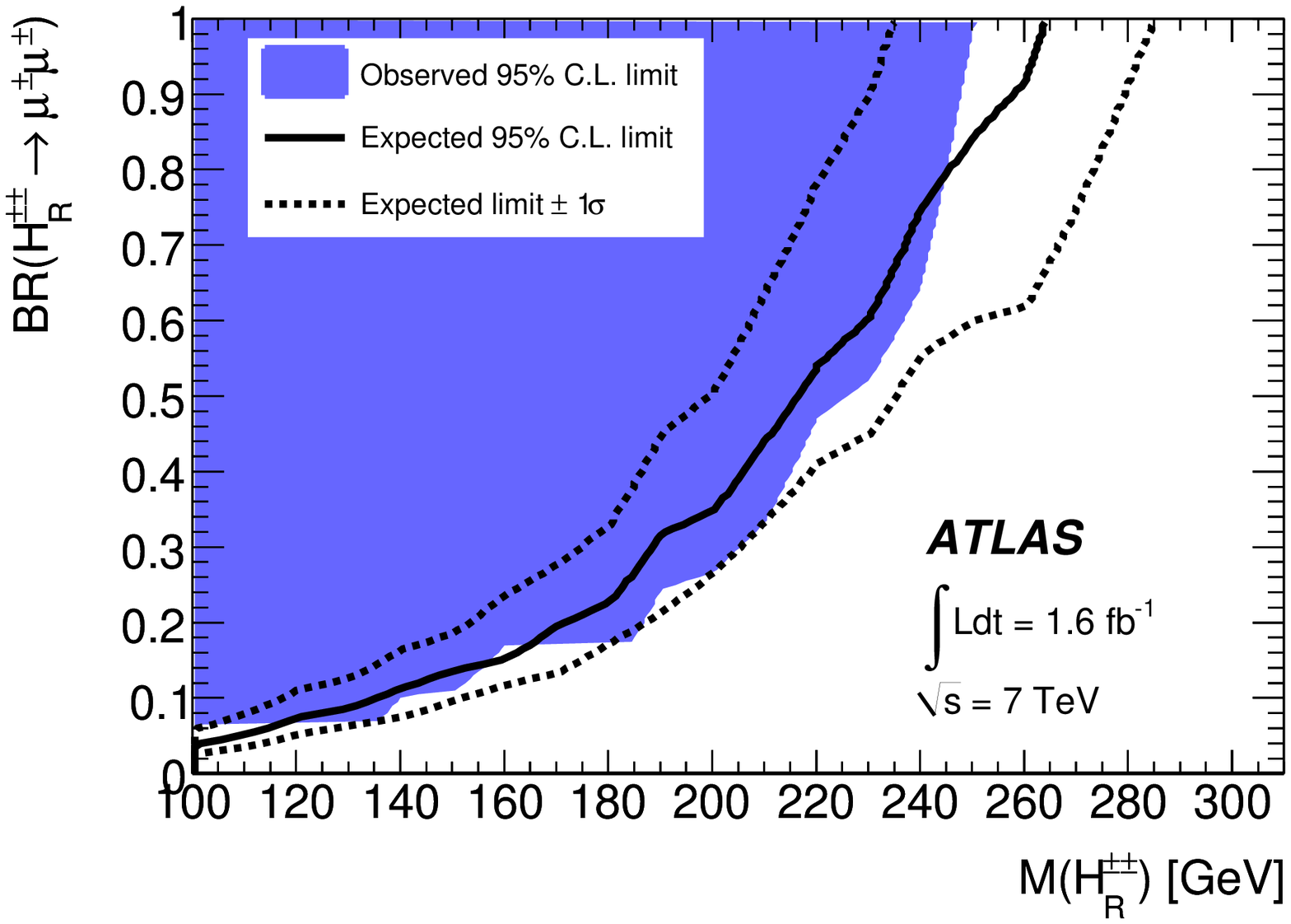}}
\caption{\label{fig:limit_vsBR} Exclusion region at 95\% C.L. of the $\hpp$ 
mass as a function of the branching ratio to muon pairs, $BR(\hpp\to\mu^\pm\mu^\pm)$, for a) 
$\hpp_L$ bosons and b) $\hpp_R$ bosons. The 
shaded areas show the observed exclusion region, the solid lines show the expected exclusion 
region, and the dashed lines show the $\pm 1\sigma$ variations of the expected exclusion region.}
\end{center}
\end{figure*}

\section{Conclusions}
\label{sec:conclusions}

An inclusive search for production of pairs of prompt like-sign muons has been presented
using a dataset corresponding to an integrated luminosity of 1.6~fb$^{-1}$
recorded with the ATLAS detector at the LHC. 
The data agree with the background expectation and no sign of new physics 
has been found. 
The data are used to place model-independent upper limits on the cross section of 
new physics processes giving rise to like-sign dimuons ranging from 5.3~fb for 
$m(\mu^\pm\mu^\pm)>300$~GeV to 58~fb for $m(\mu^\pm\mu^\pm)>15$~GeV.
In addition, constraints are placed on like-sign top-quark and doubly charged
Higgs boson production. The 95\% C.L. limit on the like-sign top-quark production cross section of 
3.7~pb is more than four times more restrictive than previous results. 
The lower mass limit on doubly charged Higgs bosons with a 100\% (33\%) branching ratio to muons is 355 (244)~GeV and 251 (209)~GeV for $\hpp$ bosons coupling to left-handed and right-handed fermions, respectively.

\section{Acknowledgements}

We thank CERN for the very successful operation of the LHC, as well as the
support staff from our institutions without whom ATLAS could not be
operated efficiently.

We also thank Michael Spira for his help with some of the theoretical
aspects of the analysis.

We acknowledge the support of ANPCyT, Argentina; YerPhI, Armenia; ARC,
Australia; BMWF, Austria; ANAS, Azerbaijan; SSTC, Belarus; CNPq and FAPESP,
Brazil; NSERC, NRC and CFI, Canada; CERN; CONICYT, Chile; CAS, MOST and
NSFC, China; COLCIENCIAS, Colombia; MSMT CR, MPO CR and VSC CR, Czech
Republic; DNRF, DNSRC and Lundbeck Foundation, Denmark; ARTEMIS, European
Union; IN2P3-CNRS, CEA-DSM/IRFU, France; GNAS, Georgia; BMBF, DFG, HGF, MPG
and AvH Foundation, Germany; GSRT, Greece; ISF, MINERVA, GIF, DIP and
Benoziyo Center, Israel; INFN, Italy; MEXT and JSPS, Japan; CNRST, Morocco;
FOM and NWO, Netherlands; RCN, Norway; MNiSW, Poland; GRICES and FCT,
Portugal; MERYS (MECTS), Romania; MES of Russia and ROSATOM, Russian
Federation; JINR; MSTD, Serbia; MSSR, Slovakia; ARRS and MVZT, Slovenia;
DST/NRF, South Africa; MICINN, Spain; SRC and Wallenberg Foundation,
Sweden; SER, SNSF and Cantons of Bern and Geneva, Switzerland; NSC, Taiwan;
TAEK, Turkey; STFC, the Royal Society and Leverhulme Trust, United Kingdom;
DOE and NSF, United States of America.

The crucial computing support from all WLCG partners is acknowledged
gratefully, in particular from CERN and the ATLAS Tier-1 facilities at
TRIUMF (Canada), NDGF (Denmark, Norway, Sweden), CC-IN2P3 (France),
KIT/GridKA (Germany), INFN-CNAF (Italy), NL-T1 (Netherlands), PIC (Spain),
ASGC (Taiwan), RAL (UK) and BNL (USA) and in the Tier-2 facilities
worldwide.

\bibliography{LikesignPaper}
\onecolumngrid

\clearpage
\onecolumngrid
\begin{flushleft}
{\Large The ATLAS Collaboration}

\bigskip

G.~Aad$^{\rm 48}$,
B.~Abbott$^{\rm 111}$,
J.~Abdallah$^{\rm 11}$,
A.A.~Abdelalim$^{\rm 49}$,
A.~Abdesselam$^{\rm 118}$,
O.~Abdinov$^{\rm 10}$,
B.~Abi$^{\rm 112}$,
M.~Abolins$^{\rm 88}$,
O.S.~AbouZeid$^{\rm 158}$,
H.~Abramowicz$^{\rm 153}$
% ...
\bigskip

% ...
$^{10}$ Institute of Physics, Azerbaijan Academy of Sciences, Baku, Azerbaijan\\
$^{11}$ Institut de F\'isica d'Altes Energies and Departament de F\'isica de la Universitat Aut\`onoma  de Barcelona and ICREA, Barcelona, Spain\\
% ...
$^{48}$ Fakult\"{a}t f\"{u}r Mathematik und Physik, Albert-Ludwigs-Universit\"{a}t, Freiburg i.Br., Germany\\
$^{49}$ Section de Physique, Universit\'e de Gen\`eve, Geneva, Switzerland\\
% ...
$^{88}$ Department of Physics and Astronomy, Michigan State University, East Lansing MI, United States of America\\
% ...
$^{111}$ Homer L. Dodge Department of Physics and Astronomy, University of Oklahoma, Norman OK, United States of America\\
$^{112}$ Department of Physics, Oklahoma State University, Stillwater OK, United States of America\\
% ...
$^{118}$ Department of Physics, Oxford University, Oxford, United Kingdom\\
% ...
$^{153}$ Raymond and Beverly Sackler School of Physics and Astronomy, Tel Aviv University, Tel Aviv, Israel\\
% ...
$^{158}$ Department of Physics, University of Toronto, Toronto ON, Canada\\
% ...
\end{flushleft}


\end{document}